\documentclass[12pt]{article}
\usepackage{graphicx}

\newsavebox{\sboxpubnumber}
\newsavebox{\sboxpubdate}
\newcommand{\pubdate}[1]{\begin{lrbox}{\sboxpubdate}{#1}\end{lrbox}}
\newcommand{\pubnumber}[1]{\begin{lrbox}{\sboxpubnumber}{\begin{tabular}{l} #1 \\
				 \usebox{\sboxpubdate}
				 \end{tabular}}
                           \end{lrbox}
                           \pubblock}
\newcommand{\Title}[1]{\begin{center} {\Large #1 } \end{center}}
\newcommand{\Author}[1]{\begin{center}{ \sc #1} \end{center}}
\newcommand{\Address}[1]{\begin{center}{ \it #1} \end{center}}
\newcommand{\andauth}{\begin{center}{and} \end{center}}
\newcommand{\pubblock}{\rightline{
			\usebox{\sboxpubnumber}}}

\newenvironment{Presented}{\begin{quotation} \begin{center}
             PRESENTED AT\end{center}\bigskip
      \begin{center}\begin{large}}{\end{large}\end{center}
      \end{quotation}}

\def\be{\begin{equation}}
\def\ee{\end{equation}}
\def\bea{\begin{eqnarray}}
\def\eea{\end{eqnarray}}

\def\apj #1 #2 #3 {ApJ, {\bf #2}, #3, (#1)}
\def\apjl #1 #2 #3 {ApJ, {\bf #2}, L#3, (#1)}
\def\apjs #1 #2 #3 {ApJS, {\bf #2}, #3, (#1)}
\def\aap  #1 #2 #3 {A\&A, {\bf #2}, #3, (#1)}
\def\mnras #1 #2 #3 {MNRAS, {\bf #2}, #3, (#1)}
\def\pra #1 #2 #3 {Phys.~Rev.~A., {\bf #2}, #3, (#1)}
\def\prb #1 #2 #3 {Phys.~Rev.~B., {\bf #2}, #3, (#1)}
\def\prc #1 #2 #3 {Phys.~Rev.~C., {\bf #2}, #3, (#1)}
\def\prd #1 #2 #3 {Phys.~Rev.~D., {\bf #2}, #3, (#1)}
\def\pre #1 #2 #3 {Phys.~Rev.~E., {\bf #2}, #3, (#1)}
\def\prl #1 #2 #3 {Phys.~Rev.~Lett., {\bf #2}, #3, (#1)}
\def\plb #1 #2 #3 {Phys.~Lett.~B., {\bf #2}, #3, (#1)}
\def\science #1 #2 #3 { Science., {\bf #2}, #3, (#1)}
\def\nature #1 #2 #3 {Nature., {\bf #2}, #3, (#1)}
\def\nphysa #1 #2 #3 {Nucl.~Phys.~A., {\bf #2}, #3, (#1)}
\def\nphysb #1 #2 #3 {Nucl.~Phys.~B., {\bf #2}, #3, (#1)}
\def\nphysbs #1 #2 #3 {Nucl.~Phys.~B.~Suppl., {\bf #2}, #3, (#1)}

\def\h#1{\hbox{${}^{#1}$H}}
\def\he#1{\hbox{${}^{#1}$He}}

\def\be#1{\hbox{${}^{#1}$Be}}

\def\omegab{\hbox{${\Omega}_{\rm b}$}}
\def\h502{\hbox{$ h^{2}_{50}$}}
\def\xinue{\hbox{$\xi_{\nu_{e}}$}}
\def\xinum{\hbox{$\xi_{\nu_{\mu}}$}}
\def\xinut{\hbox{$\xi_{\nu_{\tau}}$}}
\def\xinumt{\hbox{$\xi_{\nu_{\mu, \tau}}$}}

\def\ev{\mbox{~eV}}

\def\la{\mathrel{\mathpalette\fun <}}
\def\ga{\mathrel{\mathpalette\fun >}}
\def\fun#1#2{\lower3.6pt\vbox{\baselineskip0pt\lineskip.9pt
  \ialign{$\mathsurround=0pt#1\hfil##\hfil$\crcr#2\crcr\sim\crcr}}}

\begin{document}

\begin{titlepage}
\pubdate{\today}                    
\pubnumber{XXX-XXXXX \\ YYY-YYYYYY} 
\vfill

\Title{Universal Lepton Asymmetry: New Constraints from the Cosmic Microwave Background and Primordial Nucleosynthesis
}
\vfill

\Author{T. KAJINO, M. ORITO}

\Address{National Astronomical Observatory, Mitaka, Tokyo 181-8588, Japan;\\
E-mail: kajino@nao.ac.jp}
\vfill
\andauth
\vfill

\Author{G. J. MATHEWS}

\Address{Department of Physics and Center for Astrophysics, University of
Notre Dame, Notre Dame, IN 46556, U.S.A.}
\vfill
\andauth
\vfill

\Author{R. N. BOYD}

\Address{Department of Physics and Department of Astronomy, Ohio State
University, Columbus, OH 43210, U.S.A.}

\vfill

\begin{abstract}
We study the primordial nucleosynthesis and cosmic age 
in the presence of a net lepton asymmetry as well as baryon asymmetry.
We explore a previously unnoted
region of the parameter space in which very large baryon densities $0.1 \le \Omega_b
\le 1$ can be accommodated within the light-element constraints from
primordial nucleosynthesis.
This parameter space consists of $\nu_\mu$ and $\nu_\tau$ degeneracies
with a small $\nu_e$ degeneracy.  Constraints from cosmic microwave background 
fluctuations are also discussed \cite{orito00}.
\end{abstract}
\vfill
\begin{Presented}
    COSMO-01 \\
    Rovaniemi, Finland, \\
    August 29 -- September 4, 2001
\end{Presented}
\vfill
\end{titlepage}
\def\thefootnote{\fnsymbol{footnote}}
\setcounter{footnote}{0}

\section{Introduction}

Recent progress in cosmological deep survey has clarified progressively the origin and distribution of matter and evolution of Galaxies in the Universe.  
The origin of the light elements among them has been a topic of broad interest 
for its significance in constraining the dark matter component in the Universe
and also in seeking for the cosmological model which best fits the recent data 
of cosmic microwave background (CMB) fluctuations.
This paper is concerned with neutrinos during Big-Bang nucleosynthesis (BBN).  
In particular, we
consider new insights into the possible role which degenerate
neutrinos may have played in the early Universe.
There have been many important 
contributions toward constrainig neutrino physics. 
Hence, a discussion of neutrinos and BBN is even essential in 
particle physics as well as cosmology.

There is no observational reason to insist that the universal  lepton number
is zero. It is possible, for example,  for the individual lepton
numbers  to be large compared to the baryon number of the
Universe, while the net total lepton number is small $L \sim B$.
It has been proposed
recently \cite{casas99} that models based upon the Affleck-Dine scenario
of baryogenesis might generate naturally lepton
number asymmetry which is seven to ten orders of magnitude
larger than the baryon number
asymmetry. Neutrinos with large lepton asymmetry and masses
$\sim 0.07 \ev$ might even explain the
existence of cosmic rays  with energies in excess of
the Greisen-Zatsepin-Kuzmin cutoff \cite{gelmini99}.
It is, therefore, important
for both particle physics and cosmology to carefully scrutinize
the limits which cosmology places on the allowed range of both the lepton and
baryon asymmetries.

\section{Primordial Nucleosynthesis}

CMB power spectrum is expected to provide a precise value of
the universal baryon-mass density parameter $\Omega_b$
along with the other cosmological parameters.
It is therefore a critical test if the Big-Bang nucleosynthesis 
can predict a consistent $\Omega_b$-value.

There is a potential difficulty in the determination of $\Omega_b$
from primordial nucleosynthesis, which has been imposed by 
recent detections of a low deuterium abundance,
2.9$\times$10$^{-5}$ $\le$ D/H $\le$4.0$\times$10$^{-5}$,
in Lyman-$\alpha$ clouds along the line of sight 
to high red-shift quasars \cite{burles98}.
Primordial abundance of $^7$Li is 
constrained from the observed "Spite plateau", 
0.91$\times$10$^{-10}$ $\le$ $^7$Li/H$\le$1.91$\times$10$^{-10}$
\cite{ryan00}, 
and the $^4$He abundance by mass, 0.226$\le Y_p \le$0.247
\cite{olive99},
from the observations in the HII regions.
In order to satisfy these abundance constraints 
by a single $\Omega_b$ value, one has to assume an appreciable 
depletion in the observed abundance of $^7$Li,
which is still controversial both theoretically and observationally.

It depends on how accurately the nuclear reaction rates for
the production of $^7$Li are known.
$^7$Li abundance is strongly subject to
large error bars associated with the measured cross sections
for $^4$He($^3$H,$\gamma$)$^7$Li at $\eta \la$ 2$\times 10^{-10}$ 
and $^4$He($^3$He,$\gamma$)$^7$Be at 3$\times 10^{-10} \la \eta$.
We studied these two reactions in quantum mechanics very carefully 
and concluded that the proper 2$\sigma$ error bars could be 
1/4$\sim$1/3 of the previous ones. 
This improvement owes mostly to, first, the new precise measurement
\cite{brune94} of the cross sections for $^4$He($^3$H,$\gamma$)$^7$Li
and, second, the systematic theoretical studies \cite{kajino01}
of both reaction dynamics and quantum nuclear structures of
$^7$Li and $^7$Be, whose validity is critically
tested by electromagnetic form factors measured by
high-energy electron scattering experiments.
When our recommended error estimate is applied to the determination
of $\Omega_b$, we lose $\Omega_b$ value to explain both D/H
and $^7$Li/H simultaneously.

In order to better estimate the $\Omega_b$-value, 
we propose a new method to
determine the primordial $^7$Li by the use of isotopic abundance
ratio $^7$Li/$^6$Li in the
interstellar medium which exhibits the minimum effects
of the stellar processes including depletion effect.
Details are reported elsewhere \cite{kajino00,kawanomoto01}.

\section{Neutrino Decoupling in Lepton Asymmetric Cosmology}

Although lepton asymmetric BBN
has been studied in many papers \cite{kang92} (and references therein),
there are several differences in the present work:  
For one , we have included
finite temperature  corrections to the mass of the electron and
photon \cite{fornengo97}.
Another is that we have calculated the neutrino annihilation
rate in the cosmic comoving frame, in which the M{\o}ller velocity instead of
the relative velocity is to be used for the integration of the collision term
in the Boltzmann equations \cite{gondolo91,enqvist92}.

Neutrinos and anti-neutrinos drop out of thermal equilibrium with the
background thermal plasma  when
the weak reaction rate becomes slower than the universal expansion rate.
If the neutrinos  decouple early, they are
not heated as the particle degrees of freedom change.
Hence, the ratio of the neutrino to
photon temperatures, $T_\nu/T_\gamma$, is reduced.
The biggest drop in temperature 
occurs for a neutrino degeneracy parameter 
$\xi_\nu = \mu_\nu/T_\nu \sim 10$, where $\mu_\nu$ is the
neutrino chemical potential.
This corresponds to a decoupling
temperature above the cosmic QCD phase transition.

Non-zero lepton numbers affect nucleosynthesis in two
ways. First, neutrino degeneracy 
increases the expansion rate.
This increases the  $\he4$ production.
Secondly,
the equilibrium n/p ratio is affected by the electron neutrino chemical
potential, 
$\rm{n/p} = exp\{-(\Delta \it{M/T}_{n \leftrightarrow p}) -
\xinue\}$,
where $\Delta M$ is the neutron-proton mass difference and
$T_{n \leftrightarrow p}$ is the
freeze-out temperature for the relevant weak reactions.
This effect either increases or decreases $\he4$ production,
depending upon the sign of $\xinue$.

\begin{figure}[htb]]
    \centering
        \includegraphics[height=1.5in]{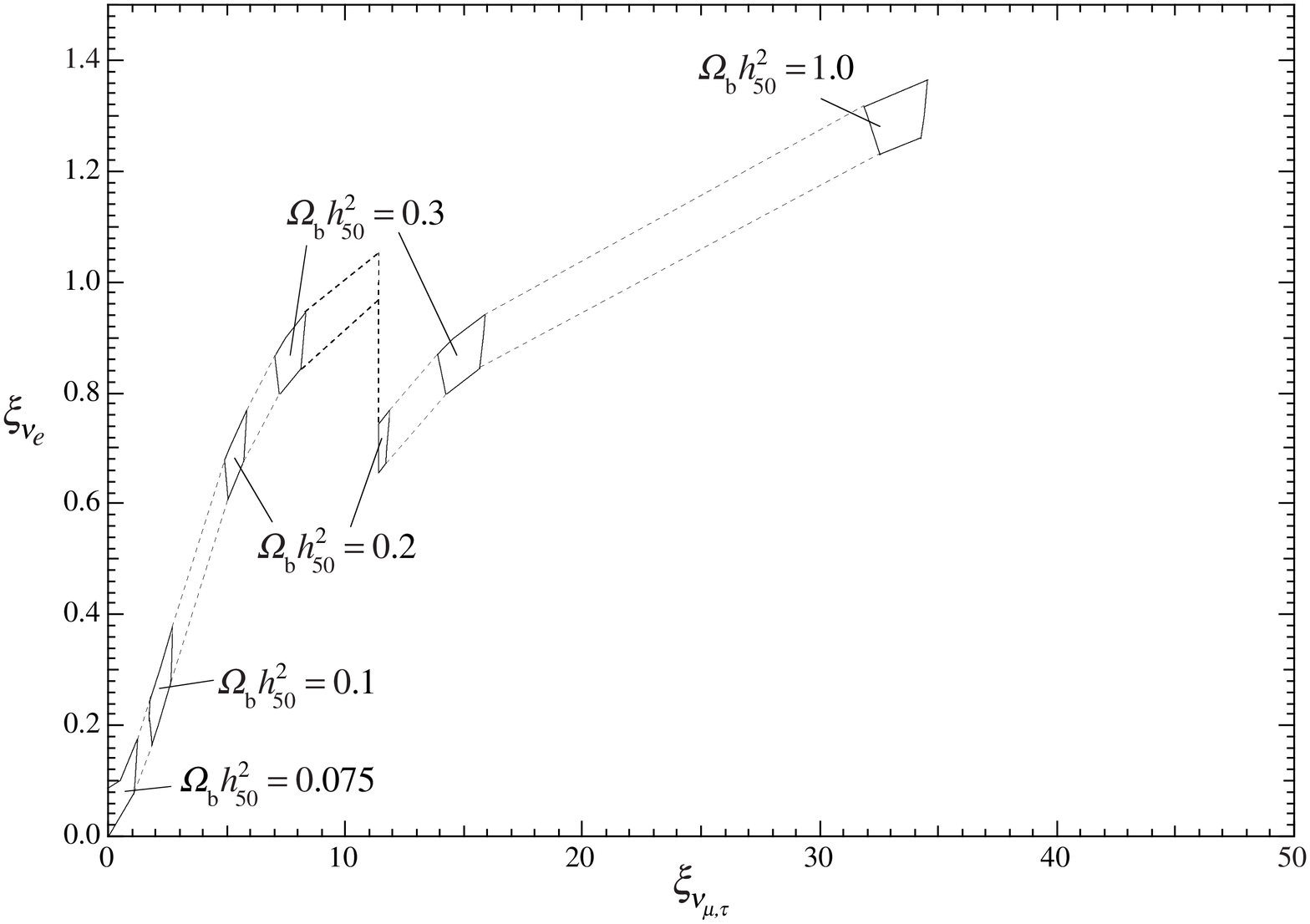}
\caption{Allowed values of $\xinue$ and $\xinumt$ for which the
constraints from light element abundances are satisfied for
values of  $\omegab \h502 =$ 0.075, 0.1, 0.2, 0.3 and 1.0 as indicated.
\label{fig1}}
\end{figure}

A third effect emphasized in this paper is that
$T_\nu/T_\gamma$ can be reduced if the neutrinos decouple
early.  This lower temperature
reduces the energy density of 
neutrinos during BBN, and slows the expansion of the
Universe. This decreases $\he4$ production.

Figure \ref{fig1} highlights the main result of this study \cite{orito00},
where we take $\xinum = -~\xinut$.
For low $\omegab \h502 $ models, only the usual low values
for $\xinue $ and $\xinumt$ are allowed.  Between
$\omegab \h502 \approx$ 0.188 and 0.3, however, more
than one allowed region emerges.
For $\omegab \h502 \ga 0.4$ only the large degeneracy solution
is allowed.  Neutrino degeneracy can even allow baryonic densities
up to $\omegab \h502 = 1$.
\begin{figure}[htb]]
        \centering
	        \includegraphics[height=1.5in]{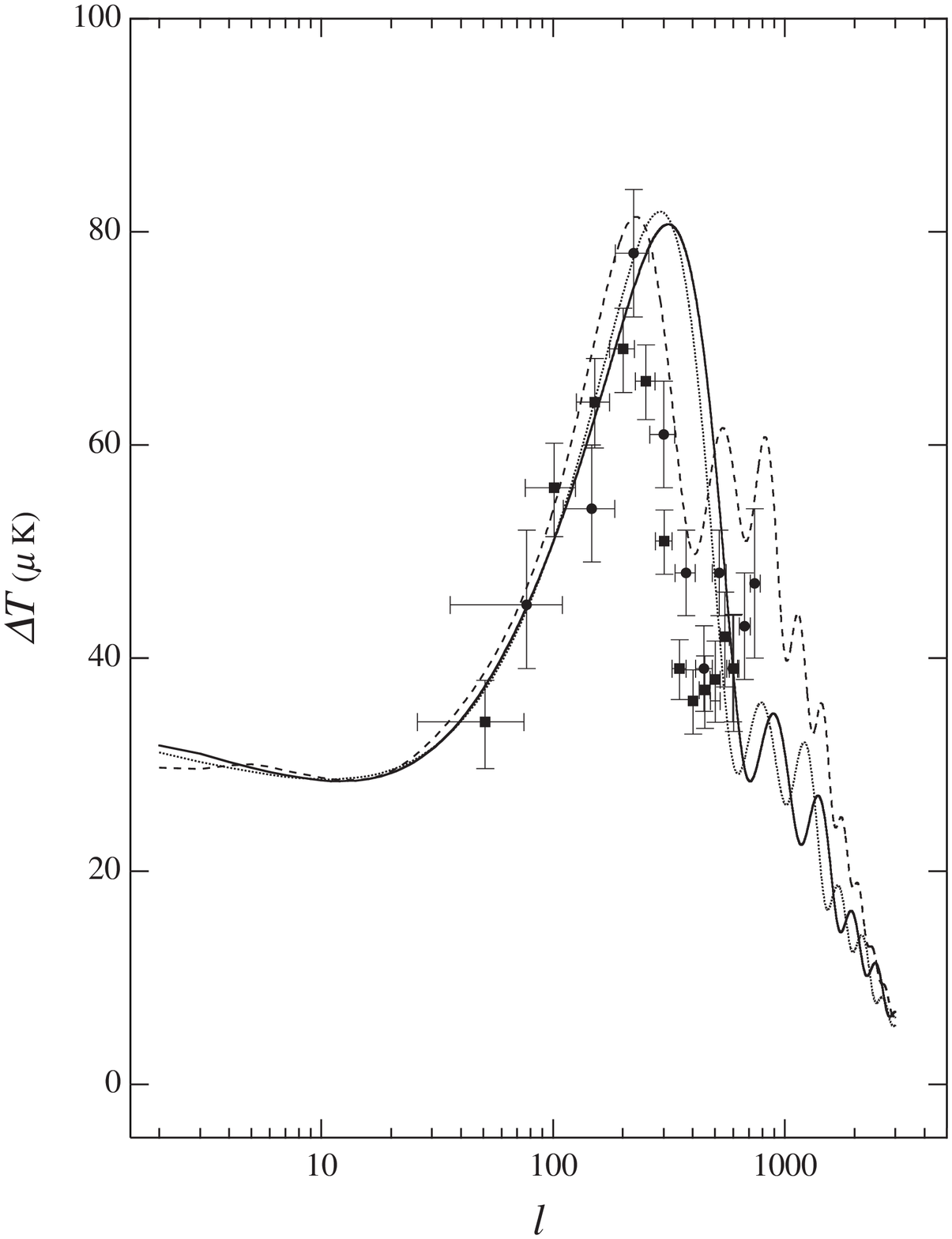}
\caption{CMB power spectrum  from 
MAXIMA-1~\protect \cite{hanany} (circles) and
BOOMERANG~\protect \cite{boomerang}
(squares) binned data compared with calculated $\Omega = 1$ models.
} 
\label{fig2}
\end{figure}

\section{Constraint from Cosmic Microwave Background}

Several recent works \cite{kinney,lesgourgues,hannestad} have shown that
neutrino degeneracy can dramatically alter the power spectrum of the CMB.
However, only small degeneracy parameters with the standard relic neutrino
temperatures have been utilized.
Here, we have calculated the CMB power spectrum to investigate effects

of a diminished relic neutrino temperature.

The solid line on Figure \ref{fig2} shows a  $\Omega_\Lambda = 0.4$
model for $n = 0.78$ which is a power law "tilt" of primordial fluctuation.  
This fit is marginally consistent with the data at a level of $5.2 \sigma$.
The dotted line in Figure \ref{fig2} shows the matter dominated
$\Omega_\Lambda = 0$ best fit model with $n = 0.83$
which is consistent with the data at
the level of $3 \sigma$.
The main differences in the fits between the large degeneracy models
and our adopted benchmark model are that the first
peak is shifted to slightly higher $l$ value
and the second peak is suppressed.
One can clearly see that the suppression of the second
acoustic peak is consistent
with our derived neutrino-degenerate models.
In particular, the MAXIMA-1 results are in very good agreement with the
predictions of our neutrino-degenerate cosmological models \cite{orito00,mathews01}.
It is clear that these new data sets substantially improve the
goodness of fit for the neutrino-degenerate models \cite{lesgourgues}.
Moreover, both data sets seem to require an increase in
the baryonic contribution to the closure density
as allowed in our neutrino-degenerate models.

\section{Cosmic Age}

There are several important implications of the neutrino degenerate Universe models.
One of them is on the cosmic age problem.
Recent baloon experiments of detecting the CMB anisotropy has
exhibited that the flat cosmology is more likely.  
Combining this with the result from high-redshift supernova search, one may deduce a finite cosmological constant $\Omega_\Lambda \sim 0.6$, leading to a cosmic age $\sim$ 15~Gy.  
If this were the case, a potential difficulty that the cosmic age is likely to be shorter than 
the age of the Milky Way might be resolved.  
However, CMB anisotropy data provide with more details of sesveral
cosmological parameters which may not necessarily accept this simplified interpretation.

In our neutrino degenerate Universe models with $\Omega = 1$,
$\Omega_\Lambda = 0.4$, and $\Omega_b h_{50}^2 = 0.1$, 
neutrino mass for $\nu_{\mu,\tau}$ is constrained 
$m_\nu \le$ 0.3 eV as far as $\Omega_\nu \le 0.5$ \cite{orito00,mathews01}.
Even should the mass be 0.3 eV, our conclusion on the primordial nucleosynthesis 
does not change at all.  Therefore, we assumed massless neutrino.
With this possible choice of the parameters in cosmology and particle physics,
we can estimate the cosmic expansion age $\approx 12 \sim 13$ Gy.
Cosmic age problem seems still remained.
Further careful studies of the age problem and also the nature of
cosmological constant \cite{yahiro01} are highly desirable.



\begin{thebibliography}{}
%
        \bibitem{orito00}    M.~Orito, T.~Kajino, G.~J.~Mathews, and R.~N.~Boyd,~
	astro-ph/0005446,~submitted to Astrophys. J. (2000)
%
        \bibitem{casas99}   A.~Casas, W.~Y.~Cheng, \& G.~Gelmini,
	\nphysb 1999 538 297
%
        \bibitem{gelmini99}    G.~Gelmini \&
        A.~Kusenko  \prl 1999 82 5202
%
	\bibitem{burles98}    Burles, S., \& Tytler, D. 1998a, Astrophys. J. 499, 699; 1998b,
	Astrophys. J. 507, 732
%
	\bibitem{ryan00}   S.~Ryan, T.~Beers, K.~Olive, B.~Fields, \& J.~Norris 2000a,
 	Astrophys. J. 530, L57;~~S.~G.~Ryan, T.~Kajino, T.~C.~Beers, T.-K.~Suzuki, D.~Romano,
 	F.~Matteucci, \& K.~Rosolankova 2000b, Astrophys. J. 549, 55
%
	\bibitem{olive99}   K.~Olive, G.~Steigman, \& T.~Walker 1999, Phys. Rep., in press
%
	\bibitem{brune94}   C.~R.~Brune, R.~W.~Kavanagh, \& C.~Rolfs 1994,

	Phys. Rev. C50, 2205
%
	\bibitem{kajino01}    T.~Kajino, M.~Orito, \& K.~Ichiki 2001, in preparation
%
	\bibitem{kajino00}    T.~Kajino, T.-K.~Suzuki, S.~Kawanomoto, \& H.~Ando 2000,
	Proc. IAU Int. Symp. No. 198, eds. L.~da~Silva, M.~Spite, \& J.~R.~de~Medeiros 
	(Astron. Soc. Pacific 2000), pp.344-349
%
	\bibitem{kawanomoto01}   S.~Kawanomoto, T.~Kajino, T.-K.~Suzuki, H.~Ando, 
	\& M.~Bessell 2001, in preparation
%
        \bibitem{kang92}  H.~Kang \&
        G.~Steigman \nphysb 1992 372 494
%
        \bibitem{fornengo97}   N.~Fornengo, C.~W.~Kim, \& J.~Song,
	\prd 1997 56 5123
%
        \bibitem{gondolo91}   P. Gondolo, \& G. Gelmini, \nphysb 1991 360 145
%
	\bibitem{enqvist92}   K. Enqvist, K, K. Kainulainen, \& V. Semikoz,
	\nphysb 1992 374 392
%
        \bibitem{kinney}   W. K. Kinney \& A. Riotto,
        \prl 1999 83 3366
%
        \bibitem{lesgourgues}   J. Lesgourgues \&
           S. Pastor,  \prd 1999 60 103521, astro-ph/0004412
%
        \bibitem{hannestad}   S. Hannestad,
        Phys. Rev. Lett., submitted, (2000), astro-ph/0005018
%
         \bibitem{boomerang}    P. Bernardls, ~et al.
         (Boomerang Collaboration) \nature 2000 404 955
%
        \bibitem{hanany}    S.~Hanany,~et al.
        (MAXIMA-1~Collaboration),~ApJL~submitted,~astro-ph/0005123
%
        \bibitem{mathews01}    G.~J.~Mathews, M.~Orito, T.~Kajino, and Y.~Wang,~
	NAOJ-Th-Ap 2001 No.9,~submitted to Phys. Rev. D. (2001)
%
        \bibitem{yahiro01}   M.~Yahiro, G.~J.~Mathews, K.~Ichiki, T.~Kajino, and 
        M.~Orito,~NAOJ-Th-Ap 2001 No.7,~submitted to Phys. Rev. D. (2001)

\end{thebibliography}
\end{document}